\documentclass[conference]{IEEEtran}
\IEEEoverridecommandlockouts
\usepackage{amsmath,amssymb,amsfonts}
\usepackage{graphicx}
\usepackage{textcomp}
\usepackage[table]{xcolor}
\usepackage{booktabs}
\usepackage{multirow}
\usepackage{array}
\usepackage{url}
\usepackage{subfig}
\usepackage{float}

\usepackage{enumitem}
\setitemize[0]{leftmargin=*}

\def\BibTeX{{\rm B\kern-.05em{\sc i\kern-.025em b}\kern-.08em
    T\kern-.1667em\lower.7ex\hbox{E}\kern-.125emX}}

\begin{document}

\title{Task-Oriented Semantic Communication for Hazard Warning and Remote Operation in Connected Vehicle Platoons
\thanks{This work was supported in part by the Swedish Knowledge Foundation (KKS) \emph{TeleDrive} project under Grant 20230076.}
}

\author{\IEEEauthorblockN{S M Sabit Bananee\textsuperscript{*}, Shahriar Hasan\textsuperscript{\dag}, Muhammad Mahbub Alam\textsuperscript{*}, Nafiul Rashid\textsuperscript{\ddag} }
\IEEEauthorblockA{\textsuperscript{*}\textit{Department of Computer Science and Engineering, Islamic University of Technology, Gazipur 1704, Bangladesh}\\
\textsuperscript{\dag}\textit{Department of Computer Science and Engineering, M\"{a}lardalen University, 721 23 V\"{a}ster\aa{}s, Sweden}\\
\textsuperscript{\ddag}\textit{Samsung Research America, Mountain View, CA 94043, USA}\\
E-mail: smsabitbananee@iut-dhaka.edu, shahriar.hasan@mdu.se, mma@iut-dhaka.edu, nafiul.rashid@gmail.com}
}

\makeatletter
\def\ps@IEEEtitlepagestyle{
  \def\@oddhead{\raisebox{0ex}[0pt][0pt]{\parbox{\textwidth}{\centering\small\itshape%
    This work has been submitted to the IEEE for possible publication. Copyright may be transferred without notice, after which this version may no longer be accessible.}}}%
  \def\@evenhead{}%
}
\makeatother

\maketitle

\begin{abstract}
Connected and automated vehicle platoons require reliable dissemination of task-relevant hazard information to enable appropriate downstream responses, with unresolved situations additionally requiring remote human intervention. However, the limited channel bandwidth of V2X communication makes high-volume sensor data transmission challenging, motivating the exchange of only task-relevant information rather than raw sensor data. This paper proposes an integrated semantic and task-oriented communication framework for cooperative hazard response and remote operation in automated vehicle platoons. The framework combines onboard instance segmentation and monocular depth estimation for task-oriented hazard decision-making with Stable Diffusion Variational Autoencoder (SD-VAE)-based semantic scene compression. Task-relevant hazard information and the compact semantic representation are jointly disseminated through a broadcast-based Emergency Semantic Message (ESM) over the existing V2X protocol stack. Evaluation demonstrates that SD-VAE consistently outperforms JPEG and JPEG 2000 on task-relevant reconstruction metrics, exhibiting graceful degradation rather than a sharp cliff effect under low-SINR conditions, demonstrating its feasibility for remote scene reconstruction when human intervention is required. Moreover, an extensive campaign of 7,290 simulation runs shows a low overall inter-vehicle collision rate of 0.99\%, with collisions occurring mainly under extreme high-speed and short-gap conditions.
\end{abstract}


\section{Introduction}
The proliferation of Connected and Automated Vehicles (CAVs) is enabling a wide range of applications aimed at making road transport safer, faster, cleaner, and more efficient. However, realizing these benefits requires connected vehicles to support a heterogeneous set of services, each with distinct requirements in terms of bandwidth, latency, and reliability. Such CAV services range from time-critical safety functions, such as cooperative collision avoidance and hazard warning, to bandwidth-intensive services, such as sensor (e.g., camera or LiDAR) data sharing for cooperative perception, high-definition map distribution, and remote operation, among others. Meeting the requirements of these diverse services over a single shared wireless channel remains a significant challenge for connected vehicle communication. Candidate vehicular communication technologies, such as IEEE 802.11p-based Dedicated Short-Range Communications (DSRC) and 5G NR-V2X sidelink operating under the UE-autonomous Mode 2 resource allocation scheme, further compound this challenge, as both provide limited channel bandwidth and impose constraints on the maximum allowable payload size that can be transmitted in a single transmission. 

Semantic communication represents a paradigm shift from conventional Shannon-based digital communication, moving beyond \emph{bit-level accuracy} toward the transmission of \emph{meaning} and \emph{task-relevant} information. This distinction was first introduced by Weaver \cite{Weaver1953Recent}, who described the general communication problem at three levels: Level A (technical), Level B (semantic), and Level C (effectiveness). Conventional vehicular communication technologies, such as DSRC and 5G NR-V2X, are primarily concerned with the \emph{Level A} problem, i.e., reliably transmitting \emph{bits} without regard to their semantic content or task relevance. Semantic communication addresses \emph{Level B} by prioritizing the \emph{meaning} conveyed by the transmitted information, whereas task- and goal-oriented communication targets \emph{Level C} by tailoring the transmitted information to the receiver's \emph{intended action}.

Semantic and task-oriented communication have recently received particular attention in vehicular communication due to stringent bandwidth, latency, and reliability constraints. For task-oriented communication, Eldeeb et al. \cite{Eldeeb2024Multi} address traffic-sign sharing over vehicle-satellite-vehicle links, Liu et al. \cite{Liu2024Adaptable} study adaptive semantic compression under varying resource constraints, and Yang et al. \cite{Yang2023Task} investigate energy-efficient cooperative relaying. For goal-oriented communication, Yildirim and Arikan \cite{Yildirim2025Goal} investigate action-based communication for autonomous vehicles, whereas Jin et al. \cite{Jin2025Goal} study semantic reconstruction under adverse channel conditions. For cooperative perception, Gan et al. \cite{Gan2026SCom} focus on bandwidth-efficient LiDAR-based 3D object detection, while Sheng et al. \cite{Sheng2024Perception} investigate importance-guided feature selection under time-varying fading channels. Furthermore, semantic-aware resource management relevant to vehicular communication has also been investigated through spectrum sharing \cite{Shao2024Spectrum} and joint user association with background-knowledge-aware bandwidth allocation \cite{Xia2024Resource}.

Despite these advances, several important gaps remain. Most existing frameworks are designed and evaluated under point-to-point communication links, disregarding the broadcast nature that is fundamental to vehicular communication \cite{Ma2024Broadcast}. Moreover, there is still limited understanding of how semantic and task-oriented communication can be realized within existing protocol stacks, particularly as semantic communication is unlikely to fully replace conventional communication in the near term \cite{chaccour2024less}. In addition, the integration of semantic and task-oriented communication into multi-agent system control loops, e.g., automated vehicle platoons, remains largely unexplored. Finally, the performance of semantic and task-oriented communication under high mobility, dense traffic, and realistic vehicular propagation conditions remains insufficiently investigated \cite{Hasan2026V2X}.

To address these gaps, this paper proposes a semantic and task-oriented communication framework for cooperative hazard response in automated vehicle platoons. When a platoon encounters a hazard, its Lead vehicle (LV) extracts the task-relevant hazard information together with a compact semantic representation of the surrounding scene and combines them into an Emergency Semantic Message (ESM) for broadcast over the existing V2X communication stack. Upon receiving the ESM, downstream platoons independently determine the required response based on the communicated hazard information and their own operating conditions, while the semantic representation can be reconstructed when additional scene information or remote assistance is required. The main contributions of this work are summarized as follows:
\begin{itemize}[leftmargin=*, topsep=2pt, itemsep=1pt, parsep=0pt, partopsep=0pt]
\item We develop an integrated semantic and task-oriented communication framework for automated vehicle platoons, combining onboard hazard detection, task-oriented decision making, semantic scene compression, and broadcast-based ESM dissemination within the existing V2X protocol stack.
\item We employ a Stable Diffusion Variational Autoencoder (SD-VAE)-based semantic encoder together with hazard-aware spatial masking and resolution-adaptive encoding to reduce the amount of scene information transmitted over the V2X channel. The proposed approach allows the semantic encoding rate to be adapted while preserving perceptual and task-relevant information under varying channel conditions.
\item We integrate the proposed framework into the PLEXE \cite{segata2014plexe} simulation environment and evaluate its semantic reconstruction, communication, and task-oriented safety performance under realistic vehicular communication scenarios.
\end{itemize}

The remainder of this paper is organized as follows. Section \ref{sec:sysArchitecture} presents the overall system architecture, while Section \ref{sec:featureExtraction} describes the end-to-end encoder-decoder pipeline, including the onboard hazard-decision process and SD-VAE-based semantic scene compression. Section \ref{sec:simScenario} describes the simulation scenario, parameter space, and performance metrics, and Section \ref{sec:Evaluation} presents the performance evaluation in terms of semantic and task-relevant reconstruction quality and task-oriented safety performance. Finally, Section \ref{sec:conclusion} concludes the paper.

\section{System Architecture} \label{sec:sysArchitecture}

\begin{figure}
\centering
\includegraphics[width=\columnwidth]{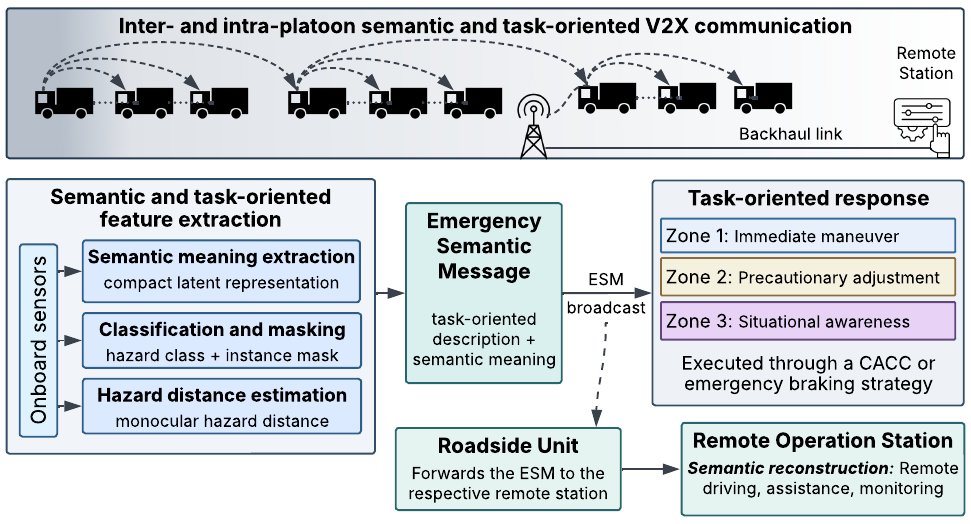}
\caption{System architecture for semantic and task-oriented hazard warning and remote operation.}
\label{fig:architecture}
\end{figure}

A schematic representation of the overall system architecture of the proposed semantic and task-oriented hazard warning and remote operation framework is depicted in Fig. \ref{fig:architecture}. As an example, Fig. \ref{fig:architecture} shows three platoons, each comprising three CAVs, driving on a highway. During cruising, the Cooperative Adaptive Cruise Control (CACC) algorithm is selected locally; as a result, the intra-platoon communication topology is dictated by the CACC algorithm adopted by each platoon. Although the focus of this work is not to study the string stability of automated vehicle platoons, the maintained gap at the time of braking, which is dictated by the control algorithm, is important for avoiding inter-vehicle collisions.

We assume that a platoon encounters a hazard of common interest, e.g., debris, unannounced road work, accidents, potholes, requiring emergency braking and dissemination of hazard messages. In such a scenario, other platoons within the immediate vicinity may also need to perform emergency braking, whereas platoons further away from the hazard may choose to perform other maneuvers, e.g., soft deceleration, lane changing, and rerouting. Moreover, if the hazard type is unknown or a CAV in a platoon cannot determine the next course of action, the vehicle can seek human intervention from a remote station according to the SAE J3016\_202104 standard \cite{sae_j3016_2021}. The semantic and task-oriented framework in Fig. \ref{fig:architecture} supports platoon cruising, hazard detection, emergency braking, as well as transmission of the camera scene from the hazard site to a remote station for remote driving, assistance, or monitoring; the framework also remains generalizable to both DSRC and 5G NR-V2X communication technologies. The constituent elements of the system architecture in Fig. \ref{fig:architecture} are outlined below.

\emph{Semantic and Task-Oriented Feature Extraction:} This is typically performed by the LV of the first platoon that encounters the hazard. Using its onboard camera, the LV performs \emph{semantic feature extraction}, \emph{hazard classification and masking}, and \emph{hazard distance estimation}, as depicted in the left-hand inset of Fig. \ref{fig:architecture}. If the hazard cannot be resolved due to an unknown class or low confidence, the semantic scene is communicated to the remote station.

\emph{Emergency Semantic Message (ESM):} The compact latent representation of the camera frame, hazard class and instance mask, and monocular hazard distance are combined to form the ESM within the payload constraints of the DSRC and 5G NR-V2X protocol stacks. The ESM contains both the task-oriented hazard information and the semantic representation of the scene and is broadcast by the affected platoon's LV.

\emph{Task-Oriented and Zone-Based Hazard Response:} Upon receiving an ESM, each platoon LV locally determines the appropriate maneuver based on its distance from the hazard, hazard type, and current state. For instance, a nearby platoon may perform emergency braking, while a platoon further away may decelerate using its CACC or seek remote assistance.

\emph{Semantic Reconstruction and Remote Operation:} A receiving platoon or remote station semantically decodes and reconstructs the transmitted scene. The reconstructed scene can support local situational awareness or, at the remote station, enable \emph{remote assistance} or \emph{remote driving}.


\begin{figure*}[!t]
\centering
\includegraphics[width=\textwidth, keepaspectratio]{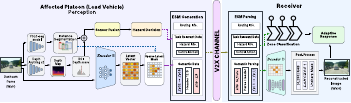}
\caption{End-to-end encoder-decoder architecture of the proposed semantic and task-oriented framework. The platoon LV extracts hazard features and sparse latent semantics, and the receiver reconstructs the scene and performs zone-based response.}
\label{fig:encoder_decoder}
\end{figure*}

\section{Semantic and Task-Oriented Encoder-Decoder Design} \label{sec:featureExtraction}

Figure \ref{fig:encoder_decoder} shows the end-to-end architecture of the proposed semantic and task-oriented communication framework, from onboard perception and hazard decision at the affected platoon to ESM transmission, zone-based adaptive response, and semantic scene reconstruction at the receiver. 

\subsection{Task-Oriented Information Extraction} \label{sec:extraction}
The onboard perception pipeline of the platoon LV transforms the captured camera frame into a compact and task-relevant representation of the surrounding scene. The LV employs YOLOv8n-seg \cite{jocher2023yolov8} and Depth Anything V2 Small \cite{yang2024depthv2} in parallel for instance segmentation and monocular depth estimation, respectively. These model outputs are subsequently fused by a lightweight, rule-based decision engine to construct an actionable state for the detected hazard.

\emph{Instance Segmentation:} The YOLOv8n-seg model is trained on the BDD100K-10K \cite{yu2020bdd100k} dataset for eight road-relevant object classes, i.e., pedestrian, rider, car, truck, bus, train, motorcycle, and bicycle. For each detection with a confidence score of $p \geq 0.35$, the model produces a bounding box, class label, confidence score, and pixel-level instance mask. From these outputs, we compute the mask-area fraction to estimate the spatial extent of the detected object and the normalized horizontal-centre fraction to determine its relevance to the ego or adjacent lane. The detected hazard region is also projected onto the semantic latent grid according to
\begin{equation}
c_x = \left\lfloor x \cdot \frac{\text{lat\_w}}{\text{orig\_w}} \right\rfloor, \qquad
c_y = \left\lfloor y \cdot \frac{\text{lat\_h}}{\text{orig\_h}} \right\rfloor,
\label{eq:latentproj}
\end{equation}
where $(x,y)$ denotes the pixel coordinate of the hazard bounding box and $(c_x,c_y)$ the corresponding latent-cell index. This projection enables the Region of Interest (ROI)-based sparse semantic encoding described in the following subsection.

\emph{Monocular Depth Estimation:} In parallel with instance segmentation, the Depth Anything V2 model is employed to estimate the relative depth of the objects detected within the same camera frame. Unlike stereo- or LiDAR-based approaches, the model produces a dense monocular depth map directly from the camera image. The resulting depth map is inverted and normalized such that pixels corresponding to objects closer to the LV approach a normalized depth score of 1.0. For each detected object, the depth values within its ROI are averaged to obtain a representative depth score. Since the subsequent decision thresholds operate on the normalized depth score rather than directly on metric depth, no explicit camera intrinsic calibration is required for this stage.

\emph{Fusion and Task-Oriented Hazard Decision:} The outputs of the instance segmentation and depth estimation models are subsequently fused to form a three-element task-oriented feature vector for every valid detection, comprising the normalized depth score, mask-area fraction, and horizontal-center fraction. A rule-based decision table with class-dependent thresholds maps this vector to one of five hazard states: \emph{Do Nothing}, \emph{Decelerate}, \emph{Lane Change}, \emph{Hard Brake}, or \emph{Cannot Resolve}. The \emph{Cannot Resolve} state is used when the LV cannot reliably determine the appropriate response, for instance, because the detected object belongs to an unknown class or the detection confidence is insufficient. When multiple objects are detected within the same frame, each detection is independently assigned a hazard state. The overall decision of the LV is subsequently selected according to the highest-priority state among all detections,
\begin{equation}
s^* = \max_i \text{priority}(s_i),
\label{eq:maxpriority}
\end{equation}
where $s_i$ represents the state assigned to detection $i$ and $\text{priority}(\cdot)$ ranks the five states from 0 for \textit{Do Nothing} to 4 for \textit{Cannot Resolve}. The resulting state $s^*$ is encoded into the ESM \texttt{controlAction} field together with the estimated hazard distance. As illustrated in Fig.~\ref{fig:encoder_decoder}, these task-relevant data are subsequently packed into the ESM together with the routing information and semantic payload before transmission over the V2X channel. The hazard distance is subsequently used by receiving platoons for zone classification, while the transmitted action represents the response selected by the affected platoon.

\subsection{Semantic Scene Compression using SD-VAE}

In parallel with the hazard decision process, the same camera frame is independently encoded using the pretrained VAE from the Stable Diffusion Architecture~\cite{rombach2022ldm}. This allows a receiving platoon or the remote operation station to reconstruct the surrounding scene when additional semantic information is required. The encoder maps an input frame to a continuous latent representation $\mathbf{z}=\mathcal{E}(\mathbf{x})\in\mathbb{R}^{4\times \frac{H}{8}\times \frac{W}{8}}$, corresponding to an eightfold spatial downscaling. The latent representation is subsequently standardized to approximately follow a $\mathcal{N}(0,1)$ distribution, which improves \texttt{float16} representation and facilitates the compression applied at the subsequent stage.

The latent representation is stored using \texttt{float16} rather than int8 precision. Although \texttt{int8} further reduces the payload size, its coarser quantization introduces smoothing artifacts that degrade the structural details required for downstream object detection and depth estimation. Therefore, \texttt{float16} is adopted as a suitable trade-off between reconstruction quality and transmission overhead. In addition, the encoder supports any input resolution divisible by~8. Consequently, the dimensions of the latent grid, and hence the resulting Bits Per Pixel (BPP), can be adjusted according to the input resolution, ranging from approximately 0.01~BPP at 64~px to 0.28~BPP at 320~px. This allows the transmitter to adapt the semantic encoding rate according to the available channel capacity and ESM payload budget without modifying the underlying architecture.

A more substantial reduction in payload size is achieved through spatial masking. At an input resolution of 320~px, the full latent representation contains up to $40\times40=1600$ spatial cells. For time-critical transmission, only the latent cells corresponding to the detected hazard region are retained, using the same pixel-to-latent projection defined in Eq. \eqref{eq:latentproj}. The retained latent cells are then losslessly compressed using zlib, which exploits the correlation among neighboring latent values to provide further reduction without additional reconstruction loss. Together, spatial masking and lossless compression reduce the payload associated with a typical hazard region by approximately 5--8 times compared with full-frame latent encoding, while preserving the scene information relevant to the detected hazard. The exact payload sizes for the considered encoding configurations, together with their corresponding reconstruction performance, are reported in detail in Section \ref{sec:Evaluation}. Furthermore, a 12-byte header is appended to each payload to specify the latent dimensions and original image resolution, allowing the receiver to decode the representation without requiring any additional signaling.

At the receiver, the ESM is first parsed to separate the task-relevant information from the semantic payload, as shown in Fig. \ref{fig:encoder_decoder}. The task-relevant information is directly used for zone classification and the corresponding adaptive response, whereas the semantic payload is decompressed and decoded when scene reconstruction is required. The reconstructed frame is then upscaled to the original image dimensions specified in the header and subjected to lightweight sharpening to restore edge details attenuated by the spatial bottleneck. For sparse hazard-ROI encoding, latent cells that are not included in the received payload are initialized to zero before decoding. The pretrained decoder can then reconstruct the surrounding background context while preserving the transmitted hazard region, thereby providing a coherent representation of the scene without requiring the full latent representation to be transmitted.

\section{Simulation Scenario and Performance Metrics} \label{sec:simScenario}
We integrate the proposed semantic and task-oriented framework in Fig.~\ref{fig:encoder_decoder} into the PLEXE simulation framework \cite{segata2014plexe}. PLEXE is built on top of the VANET simulator Veins \cite{sommer2011veins}, which bidirectionally couples OMNeT++ and SUMO and implements the IEEE 802.11p protocol stack.

We consider a multi-platoon highway scenario comprising four platoons cruising across two lanes, while a third lane is occupied by non-platooning CAVs that generate periodic beacons at 10 Hz. These additional CAVs are introduced to create background communication traffic and, consequently, a more realistic level of channel contention and interference. When the LV of the first platoon encounters an imaginary hazard at a predefined time, it starts its emergency braking maneuver, constructs the ESM as per Fig. \ref{fig:encoder_decoder}, and broadcasts the ESM. Each platoon LV in the downstream direction relays this ESM and starts the emergency braking maneuver following their respective control algorithms. To account for multipath propagation and path-loss effects, we employ the Nakagami-$m$ fading model with $\alpha=3$ together with the free-space path-loss model. Moreover, to represent the remote operation scenario, an RSU is introduced into the simulation, which can receive the transmitted ESM and reconstruct the corresponding scene from the received semantic representation. Similarly, the following platoons can also reconstruct the scene from the received ESM.

\begin{table}[t]
\caption{Simulation parameter space ($7{,}290$ runs).}
\label{tab:simparams}
\centering
\resizebox{\columnwidth}{!}{%
\begin{tabular}{
>{\raggedright\arraybackslash}p{2.8cm}
>{\raggedright\arraybackslash}p{4.5cm}
>{\centering\arraybackslash}p{0.8cm}
}
\toprule
\rowcolor[gray]{0.88} \textbf{Parameter} & \textbf{Values} & \textbf{Count} \\
\midrule
\rowcolor[gray]{0.95} Platoon size        & 4, 6, 8 vehicles                                & 3 \\
non-platooning vehicles      & 50, 100, 200                                    & 3 \\
\rowcolor[gray]{0.95} Leader speed        & 80, 100, 120 km/h                               & 3 \\
CACC algorithm      & Rajamani et al. \cite{Rajamani2000CACC}, Ploeg et al. \cite{ploeg2011design}                                     & 2 \\
\rowcolor[gray]{0.95} Inter-platoon gap   & 50, 100, 150 m                                  & 3 \\
Deceleration rate   & $-5$, $-6$, $-8$ m/s$^{2}$                     & 3 \\
\rowcolor[gray]{0.95} Repetitions         & 3 per combination                               & 3 \\
Hazard action       & \emph{Do Nothing}, \emph{Lane Change}, \emph{Decelerate}, \emph{Hard Brake}, \emph{Cannot Resolve}          & 5  \\
\midrule
\rowcolor[gray]{0.95} \textbf{Total}      &  & \textbf{7{,}290} \\
\bottomrule
\end{tabular}
}
\end{table}

An extensive simulation campaign comprising 7,290 runs is performed by varying the parameters listed in Table~\ref{tab:simparams}. For platoon cruising, we employ the CACC controllers proposed by Rajamani et al.~\cite{Rajamani2000CACC} and Ploeg et al.~\cite{ploeg2011design}. For the Rajamani CACC, an inter-vehicle gap of 5~m is considered, whereas the Ploeg CACC operates with a time headway of 0.5~s. Other PHY-layer parameters, e.g., transmission power, bit rate, and noise floor, follow the IEEE 802.11p configuration. For the hazard action parameter in Table~\ref{tab:simparams}, emergency braking for both the \emph{decelerate} and \emph{hard-brake} actions is performed at deceleration rates of $-5$, $-6$, or $-8$~m/s$^2$; milder rates were found not to affect collision occurrence and are therefore omitted for conciseness. Nevertheless, deceleration at lower rates can have an impact on string stability, the assessment of which remains outside the scope of this paper. Further, for the \emph{cannot resolve} action, we consider a scenario where the platoon brakes at $-5$, $-6$, or $-8$~m/s$^2$ to reduce its speed to 20~km/h from its initial speed.

To evaluate the reconstructed scene, we employ four complementary metrics. Structural Similarity Index Measure (SSIM) and Learned Perceptual Image Patch Similarity (LPIPS) assess pixel-level structure and perceptual similarity, respectively. To measure task-relevant reconstruction quality, we additionally use mean Intersection over Union (mIoU) and mean Average Precision at an IoU threshold of 0.5 (mAP@0.5), which evaluate the preservation of semantic information for downstream segmentation and object detection. To assess platoon safety under the proposed task-oriented hazard response, we record inter-vehicle collision occurrences across the 7,290 simulation runs and perform a decision-tree-based root-cause analysis, detailed in Subsection \ref{sec:taskPerformance}, to identify the parameter combinations associated with collisions. 

\section{Performance Evaluation} \label{sec:Evaluation}
This section evaluates the reconstruction quality and task-oriented safety performance of the proposed framework under the simulation scenario described above.
\subsection{Semantic and Task-Relevant Reconstruction Quality}
\begin{figure}[t]
\centerline{\includegraphics[width=\columnwidth]{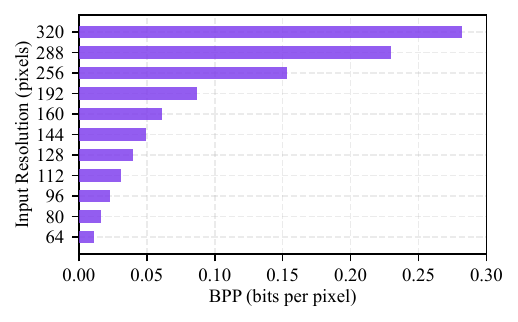}}
\caption{SD-VAE BPP as a function of input resolution, illustrating continuous rate control via resolution-adaptive encoding.}
\label{fig:bpp_size}
\end{figure}

The resolution-adaptive encoding of SD-VAE allows the transmission rate to be controlled through the selected input resolution. As shown in Fig. \ref{fig:bpp_size}, the required Bits Per Pixel (BPP) gradually increases from approximately 0.01 at 64 px to 0.28 at 320 px. This provides flexibility for selecting the semantic encoding rate according to the available communication resources and the required reconstruction quality.

\begin{figure}[!t]
\centering
\includegraphics[width=0.49\columnwidth]{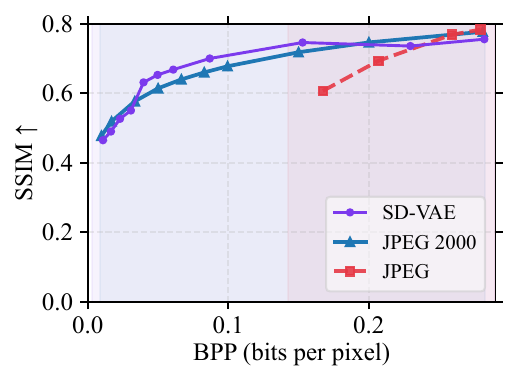}
\hfil
\includegraphics[width=0.49\columnwidth]{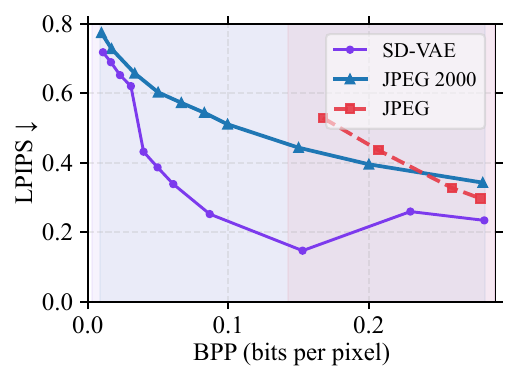}
\\[-0.6ex]
\includegraphics[width=0.49\columnwidth]{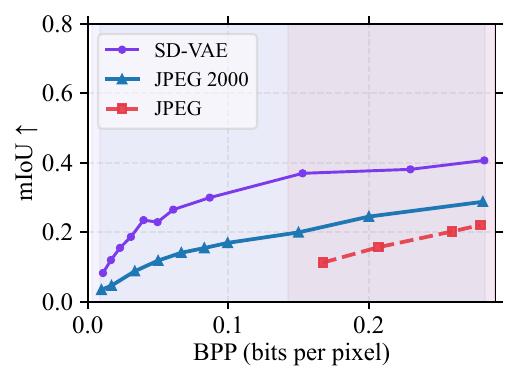}
\hfil
\includegraphics[width=0.49\columnwidth]{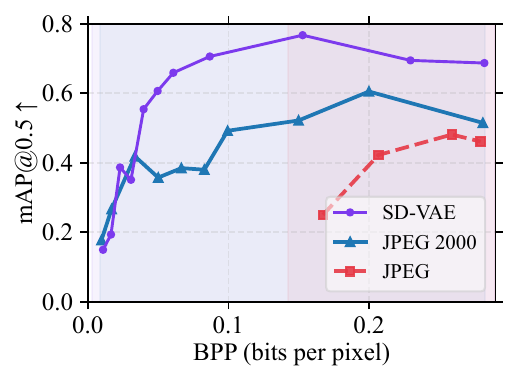}
\caption{Rate-distortion performance of SD-VAE compared against JPEG and JPEG~2000 across pixel-level (\emph{SSIM, top left}), perceptual (\emph{LPIPS, top right}), and task-relevant (\emph{mIoU, bottom left}; \emph{mAP@0.5, bottom right}) reconstruction quality as a function of BPP.}
\label{fig:bpp_all}
\end{figure}

Fig. \ref{fig:bpp_all} compares SD-VAE against the JPEG and JPEG 2000 baselines for the same BPP range considered in Fig. \ref{fig:bpp_size} using four metrics. In Fig. \ref{fig:bpp_all}, notice that JPEG does not have a comparable operating point below 0.167 BPP, as its block-based Discrete Cosine Transform (DCT) coding lacks the adaptability needed for scalable low-rate encoding. Among the four considered metrics, the advantage of SD-VAE is most evident for the two task-relevant metrics. SD-VAE reaches an mAP@0.5 of approximately 0.66 at only 0.06 BPP, which already exceeds the best value achieved by JPEG 2000 over the entire considered range, i.e., approximately 0.60 at 0.20\,BPP. A similar trend can be observed for mIoU, where SD-VAE reaches approximately 0.30 at 0.09 BPP, while JPEG 2000 does not reach this value even at its highest tested rate of approximately 0.28 BPP. For LPIPS, SD-VAE reaches its best perceptual score of approximately 0.15 at 0.15 BPP, which is significantly lower than JPEG 2000 within the considered range. In terms of SSIM, SD-VAE performs better at low-to-moderate rates, whereas at the highest tested BPP, JPEG and JPEG 2000 slightly outperform SD-VAE. This is mainly due to the pixel-level nature of SSIM, whereas the higher mIoU and mAP@0.5 values with SD-VAE indicate better preservation of the information required for downstream segmentation and object detection, which is the main goal of semantic and task-oriented communication.

\begin{figure}[!t]
\centering
\includegraphics[width=0.49\columnwidth]{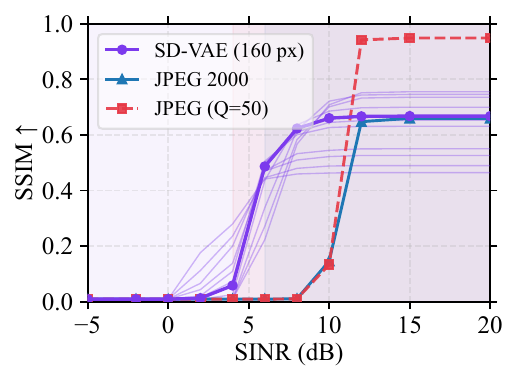}
\hfil
\includegraphics[width=0.49\columnwidth]{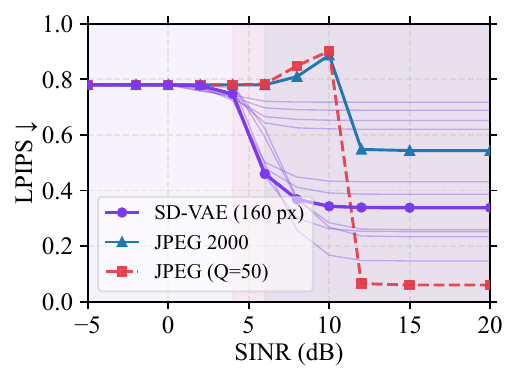}
\\[-0.6ex]
\includegraphics[width=0.49\columnwidth]{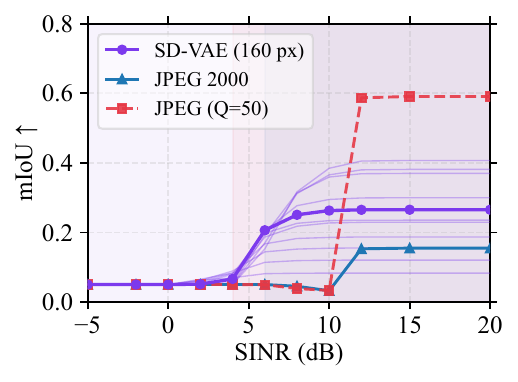}
\hfil
\includegraphics[width=0.49\columnwidth]{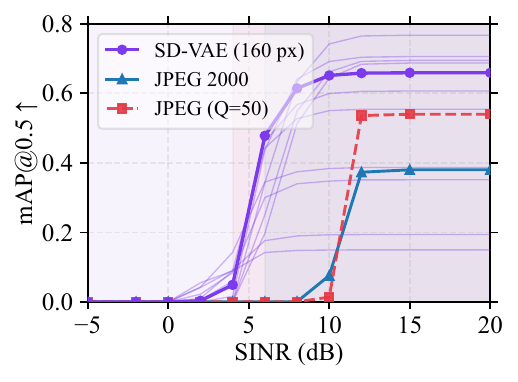}
\caption{Decoded scene quality of SD-VAE, evaluated across all tested input resolutions (thin lines, 64--320\,px, with 160\,px highlighted in bold), compared to JPEG and JPEG~2000 for varying SINR (dB).}
\label{fig:sinr_all}
\end{figure}

Fig. \ref{fig:sinr_all} presents the same four metrics as a function of SINR. JPEG and JPEG 2000 are evaluated at fixed operating points, while the thin purple curves represent the different SD-VAE input resolutions from 64 to 320 px, and the bold curve highlights the 160 px operating point used for direct comparison. An important observation from Fig. \ref{fig:sinr_all} is that the JPEG and JPEG 2000 baselines exhibit the cliff effect, i.e., their performance remains close to the minimum level over most of the low-SINR region and then improves sharply between approximately 10 and 12 dB. In contrast, SD-VAE begins to recover from around 4--6 dB and improves more gradually before reaching its plateau at approximately 10--12 dB. This behavior can also be observed consistently across the different SD-VAE input resolutions. The ability of SD-VAE to preserve useful perceptual and task-relevant information under low-to-moderate SINR conditions is particularly important for vehicular communication, where the channel condition may deteriorate rapidly due to high-speed mobility, dynamic topology changes, and variations in traffic conditions.

\subsection{Task-Oriented Safety Performance} \label{sec:taskPerformance}
Table~\ref{tab:yolo} reports the validation performance of YOLOv8n-seg on BDD100K-10K after fine-tuning for 100 epochs on an RTX 3090 GPU. The detection performance varies considerably across classes; for instance, car, the most frequently represented class, achieves the highest Box mAP50 of 0.673, whereas less represented classes such as bicycle and rider achieve much lower values of 0.094 and 0.145, respectively. Recall is also consistently lower than precision across almost all classes, with overall values of 0.268 and 0.655, indicating that the model produces fewer false detections but misses a larger fraction of actual instances, particularly for the less represented classes. The train class reports a precision of 1.0 with a recall of 0, which is attributable to its very limited representation in the validation set.

\begin{table}[t]
\caption{\small YOLOv8n-seg validation results on BDD100K-10K.}
\label{tab:yolo}
\centering
\resizebox{\columnwidth}{!}{%
\begin{tabular}{
>{\raggedright\arraybackslash}p{1.1cm}
>{\centering\arraybackslash}p{1.2cm}
>{\centering\arraybackslash}p{1.2cm}
>{\centering\arraybackslash}p{1.2cm}
>{\centering\arraybackslash}p{1.2cm}
>{\centering\arraybackslash}p{1.2cm}
}
\toprule
\rowcolor[gray]{0.88} \textbf{Class} & \textbf{Box Precision} & \textbf{Box Recall} & \textbf{Box F1 Score} & \textbf{Box mAP50} & \textbf{Mask mAP50} \\
\midrule
\rowcolor[gray]{0.95} Pedestrian & 0.756 & 0.419 & 0.539 & 0.500 & 0.464 \\
Car        & 0.784 & 0.623 & 0.694 & 0.673 & 0.611 \\
\rowcolor[gray]{0.95} Truck      & 0.603 & 0.343 & 0.437 & 0.409 & 0.384 \\
Bus        & 0.702 & 0.311 & 0.431 & 0.363 & 0.360 \\
\rowcolor[gray]{0.95} Motorcycle & 0.638 & 0.255 & 0.364 & 0.325 & 0.324 \\
Bicycle    & 0.397 & 0.062 & 0.107 & 0.094 & 0.079 \\
\rowcolor[gray]{0.95} Rider      & 0.359 & 0.133 & 0.194 & 0.145 & 0.102 \\
Train      & 1.000 & 0.000 & 0.000 & 0.006 & 0.006 \\
\midrule
\rowcolor[gray]{0.95} \textbf{All} & \textbf{0.655} & \textbf{0.268} & \textbf{0.380} & \textbf{0.314} & \textbf{0.291} \\
\bottomrule
\end{tabular}
}
\vspace{-3mm}
\end{table}

\begin{figure}[!t]
\centerline{\includegraphics[width=\columnwidth]{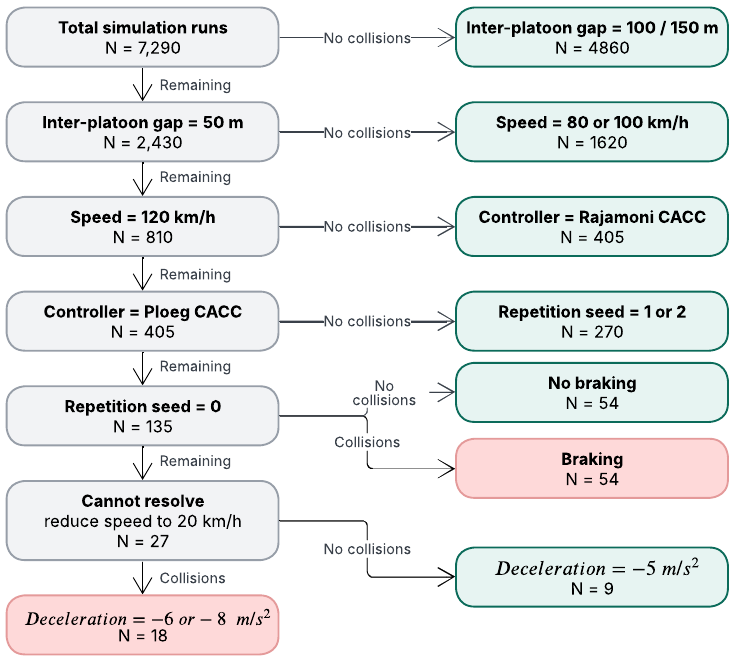}}
\caption{Decision path for collision occurrence across all 7,290 simulation runs, identified by a decision tree classifier. Every leaf reaches exactly 0\% or 100\% collision rate.}
\label{fig:collision_tree}
\end{figure}

A total of 72 collisions occurred across the 7,290 simulation runs, corresponding to an overall collision rate of 0.99\%. In order to identify the parameter combinations in Table \ref{tab:simparams} that dictate the collision occurrence, we trained a decision tree classifier (\texttt{scikit-learn}, Gini criterion, unrestricted depth) on the results of the 7,290 simulation runs, using the eight parameters in Table \ref{tab:simparams} as input features and collision occurrence as the binary target. The classifier achieves 100\% accuracy, and Fig. \ref{fig:collision_tree} shows the resulting decision path.

The inter-vehicle collisions mainly occur when the inter-platoon gap is 50 m, the leader speed is 120 km/h, and the Ploeg CACC is used. Fig. \ref{fig:collision_tree} further shows that when the platoons reduce the speed to 20 km/h due to a \emph{Cannot resolve} action at a deceleration rate $-6$ or $-8$ \,m/s\textsuperscript{2}, there are 18 collision cases. In addition, the largest number of collisions (i.e., 54) is due to the emergency braking maneuver. Importantly, the observed collisions are not attributable to communication outages. All 72 collision-positive runs occurred at an Emergency Semantic Message-Delivery Ratio (ESM-DR) of 1.0, confirming that the collisions arise from the vehicle-following response to a successfully delivered task-relevant command rather than from packet loss. Therefore, the collisions can likely be attributed to an inter-platoon collision occurring at the 50 m gap and 120 km/h speed with the Ploeg CACC. The leading vehicle of the following platoon begins braking upon receiving the ESM broadcast from the leading vehicle of the platoon ahead. At high speed, however, the braking distance required by the leading platoon increases substantially, leaving insufficient room within the 50 m inter-platoon gap for the following platoon to respond to the ESM in time.

\section{Conclusion} \label{sec:conclusion}
This paper proposed an integrated semantic and task-oriented communication framework for cooperative hazard response and remote operation in automated vehicle platoons. The framework combines onboard hazard detection and task-oriented decision-making with compact semantic scene transmission through broadcast-based ESM dissemination over the existing V2X protocol stack. Simulation results show that the SD-VAE encoder supports flexible selection of the semantic encoding rate to balance available communication resources and the desired reconstruction quality. It consistently outperforms the conventional JPEG and JPEG 2000 baselines on task-relevant reconstruction metrics, reaching an mAP@0.5 of 0.66 at only 0.06 bits per pixel, while exhibiting graceful degradation under low-SINR conditions instead of the sharp cliff effect observed for the conventional baselines. These results demonstrate its potential for supporting remote operation of connected vehicles. Furthermore, across 7,290 simulation runs, the proposed hazard-aware, task-oriented ESM broadcast results in a low overall inter-vehicle collision rate of 0.99\%, with collisions occurring only under extreme conditions, i.e., at 120 km/h with a short inter-platoon gap. Future work will investigate lightweight and computationally efficient models for real-time deployment on vehicular platforms.

\bibliographystyle{IEEEtran}
\bibliography{references}

\end{document}